\documentclass[lettersize,journal]{IEEEtran}
\usepackage{amsmath,amsfonts}
\usepackage{algorithmic}
\usepackage{algorithm}
\usepackage{array}
\usepackage[caption=false,font=normalsize,labelfont=sf,textfont=sf]{subfig}
\usepackage{textcomp}
\usepackage{stfloats}
\usepackage{url}
\usepackage{verbatim}
\usepackage{graphicx}
\usepackage{cite}
\usepackage[
	colorlinks=true,
	urlcolor=blue,
	linkcolor=black,
	citecolor=black
]{hyperref}
\begin{document} 
	\title{ST-Topo GAN: A Motor EEG-to-EMG Decoding Model Matched to Wrist Movement Complexity} 
	\author{ Ye Sun\textsuperscript{1,2,3,\textdagger}, Mingxuan Qu\textsuperscript{1,3,\textdagger}, Jing Wang\textsuperscript{4}, Dezhong Yao\textsuperscript{1,3,5,6}, and Gang Liu\textsuperscript{1,2,3,*} \thanks{\textsuperscript{1}School of Electrical and Information Engineering, Zhengzhou University, Zhengzhou 450001, China. \textsuperscript{2}HumVerse (Zhengzhou) Technologies Co., Ltd., Zhengzhou 450001, China. \textsuperscript{3}Henan Provincial Key Laboratory of Brain Science and Brain-Computer Interface Technology, Henan, China. \textsuperscript{4}School of Mechanical Engineering, Xi'an Jiaotong University, Xi'an 710049, China. \textsuperscript{5}Clinical Hospital of Chengdu Brain Science Institute, MOE Key Laboratory for NeuroInformation, Brain-Apparatus Communication Institute, University of Electronic Science and Technology of China, Chengdu, China. \textsuperscript{6}Research Unit of NeuroInformation, 2019RU035, Chinese Academy of Medical Sciences, Chengdu 611731, China.} \thanks{\textsuperscript{\textdagger}Ye Sun and Mingxuan Qu contributed equally to this work.} \thanks{The implementation of ST-Topo GAN is publicly available on GitHub at \protect\url{https://github.com/liugang1234567/st_topo_gan}.} \thanks{\textsuperscript{*}Corresponding author: Gang Liu (email: gangliu\_@zzu.edu.cn).} \thanks{This work was supported by the National Natural Science Foundation of China under Grants 62303423 and 62373295; the STI 2030--Major Project under Grant 2022ZD0208500; the China Postdoctoral Science Foundation under Grants 2024T170844 and 2023M733245; and the Henan Province Key Research and Development and Promotion Special Project under Grant 252102311096.} } 
	\maketitle

\begin{abstract} 

	The wrist plays a critical role in upper-limb function by enabling precise hand positioning, force regulation, and object manipulation. Continuous brain--muscle interfaces (BMIs) offer a promising approach for motor restoration by decoding neural activity into muscle activation signals. However, existing EEG-to-EMG models have mainly been developed for tasks with relatively stable muscle synergies and may be less effective for the heterogeneous and weakly coupled neuromuscular organisation involved in wrist movements. This paper proposes ST-Topo GAN, a Spatial--Temporal Topological Generative Adversarial Network for continuous EEG-to-EMG decoding of wrist movements. The framework integrates multi-band EEG representation, sensorimotor cortical topology modelling, and conditional adversarial learning to reconstruct multi-channel iEMG activation. The model was evaluated through cross-task comparison, wrist EEG-to-iEMG decoding, and ablation experiments. Compared with the WAY-EEG-GAL grasp-and-lift dataset, the wrist dataset exhibited lower inter-muscle activation similarity and greater decoding difficulty for conventional models. ST-Topo GAN achieved an average PCC of 0.4436 on the wrist dataset, outperforming all evaluated baselines, while the ablation study confirmed the contribution of its key components. These results support the effectiveness of ST-Topo GAN for continuous wrist EEG-to-iEMG decoding.
	
\end{abstract}


\section{Introduction}

The wrist is essential for upper-limb function, enabling precise hand positioning, force regulation, and object manipulation during daily activities such as grasping, writing, and tool use~\cite{ref3,ref4}. Impaired wrist control reduces hand dexterity and functional independence, particularly after neurological disorders such as stroke and spinal cord injury. Restoring continuous wrist movement is therefore an important goal of upper-limb neurorehabilitation~\cite{ref1,ref2}.

Continuous brain--muscle interfaces (BMIs) offer a promising approach to motor restoration by translating cortical activity into continuous muscle activation signals~\cite{ref5}. Compared with discrete movement classification, continuous EEG-to-EMG decoding captures the temporal evolution and activation intensity of individual muscles, providing richer information for adaptive motor control and rehabilitation~\cite{ref6,ref7}. EEG is particularly suitable for BMI applications because it is non-invasive and offers high temporal resolution~\cite{ref8,ref9}. Previous studies have demonstrated the feasibility of reconstructing continuous EMG activity from EEG signals~\cite{ref10,ref11}.

Recent advances in deep learning have improved EEG-to-EMG decoding by learning nonlinear representations from multichannel neural signals. CNN-, recurrent-, and Transformer-based architectures have been applied to capture spatial and temporal characteristics of motor EEG activity~\cite{ref12,ref13,ref14}. The WAY-EEG-GAL dataset, which contains synchronised EEG and EMG recordings during grasp-and-lift movements, has been widely used as a benchmark for continuous neural-to-muscular decoding~\cite{ref15,ref16,ref17}. However, existing approaches have been mainly evaluated on tasks with relatively consistent and strongly correlated muscle activation, leaving decoding under more heterogeneous and weakly correlated muscle activity, such as wrist movements, less explored.

Wrist flexion and extension require coordinated recruitment of multiple forearm muscles with distinct agonistic, antagonistic, and stabilising roles~\cite{ref18,ref19,ref20}. Compared with grasp-and-lift movements, wrist control involves more independent modulation of individual muscles according to movement direction, activation intensity, and joint stability. Accordingly, wrist movements may exhibit weaker inter-muscle correlations and more independent activation patterns~\cite{ref21,ref22,ref23}, placing greater demands on modelling the mapping between distributed cortical activity and heterogeneous muscle responses.

These characteristics also impose greater requirements on EEG representation. Wrist-related cortical activity involves multiple oscillatory components associated with motor preparation, execution, and sensorimotor feedback, while motor activity is distributed across neighbouring sensorimotor cortical regions. Treating EEG electrodes as independent temporal sequences may therefore discard relevant spatial information. In addition, cortical activity and muscle recruitment are related through nonlinear and temporally extended processes involving neural preparation, muscle activation, and sensory feedback~\cite{ref24,ref25}.

Existing EEG-to-EMG approaches do not fully account for these requirements. Many models do not explicitly preserve the spatial organisation of sensorimotor cortical activity, while point-wise regression objectives may produce over-smoothed muscle activation. These limitations are particularly relevant for wrist movements involving heterogeneous and weakly coupled muscle activation patterns. An effective wrist EEG-to-iEMG decoder therefore requires representations that capture frequency-specific cortical activity, spatial organisation, and physiologically meaningful muscle activation.

\begin{figure}[t]
	\centering
	\includegraphics[width=0.5\textwidth]{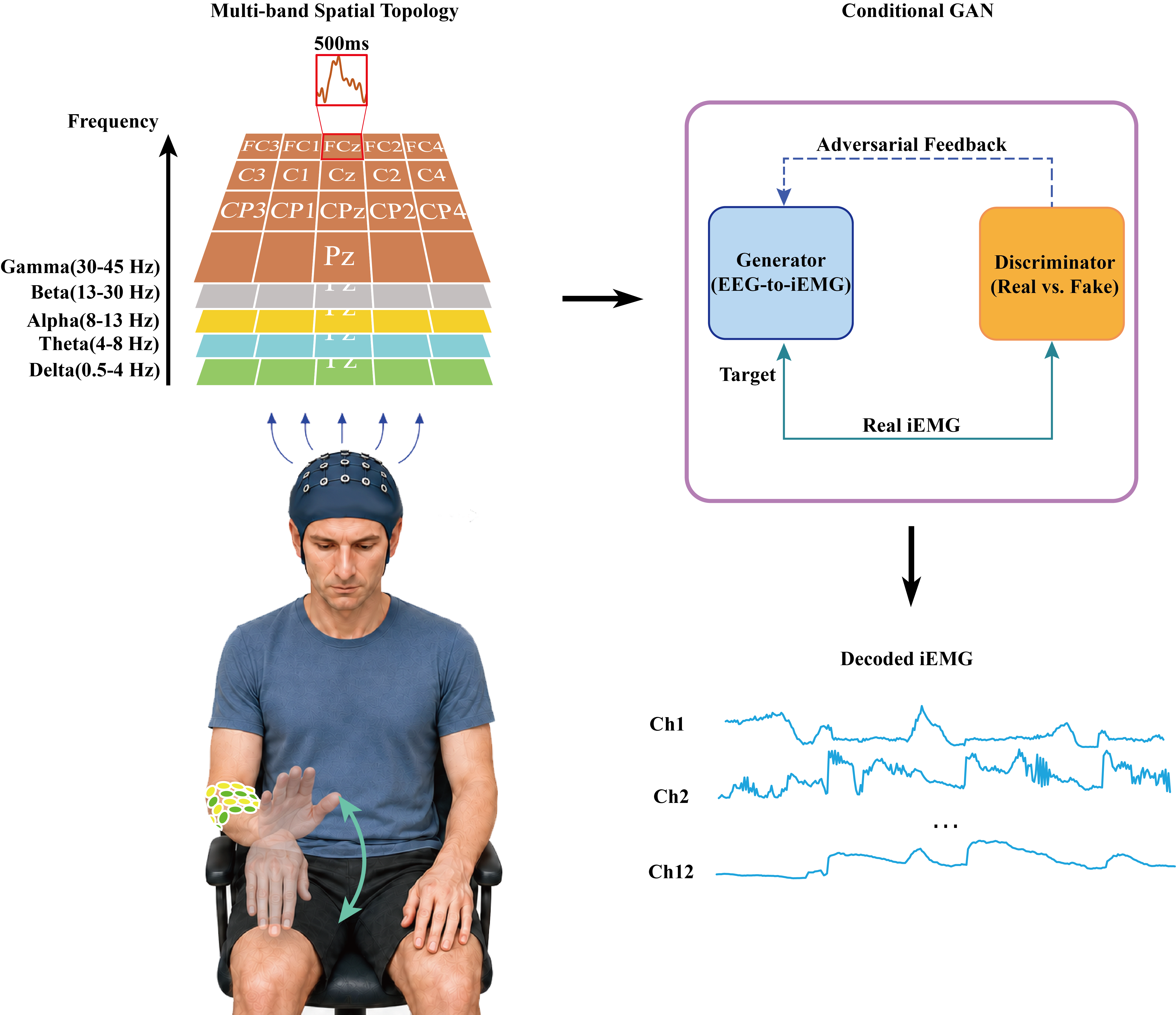}
	\caption{Overview of the proposed ST-Topo GAN framework for continuous wrist EEG-to-iEMG decoding. The proposed framework reconstructs continuous multi-channel iEMG signals through conditional adversarial learning, where generated and real iEMG patterns are compared to improve the physiological plausibility of decoded muscle activation.}
	\label{fig:overview}
\end{figure}

To address these challenges, this paper proposes a Spatial--Temporal Topological Generative Adversarial Network (ST-Topo GAN) for continuous wrist EEG-to-iEMG decoding, as illustrated in Fig.~\ref{fig:overview}. The proposed framework decomposes EEG into multiple frequency bands and projects them onto a sensorimotor cortical topology to preserve spectral and spatial characteristics. Conditional adversarial learning is further introduced to reconstruct continuous multi-channel iEMG while constraining the generated muscle activation to physiologically plausible distributions.

By integrating frequency-specific spatial topology, temporal modelling, and conditional adversarial learning, ST-Topo GAN captures distributed sensorimotor activity and constrains the generated iEMG towards physiologically plausible activation patterns. The framework was evaluated through cross-task comparison with the WAY-EEG-GAL dataset, continuous wrist EEG-to-iEMG decoding against representative deep learning models, and ablation analysis of its key components.

The main contributions of this work are summarised as follows:

\begin{enumerate}
	
	\item To the best of our knowledge, this is the first study to investigate continuous multi-channel iEMG decoding from scalp EEG during dynamic wrist flexion and extension. Cross-task experiments show that conventional models performing effectively on grasp-and-lift movements with relatively high inter-muscle activation similarity exhibit consistent performance degradation on wrist movements characterised by lower and more heterogeneous inter-muscle correlations.
	
	\item This paper proposes ST-Topo GAN, a continuous motor EEG-to-iEMG decoding framework specifically designed for wrist movements. The framework integrates multi-band cortical representation, wrist-related spatial topology modelling, and conditional adversarial generation to preserve frequency-specific and spatial sensorimotor information while reconstructing heterogeneous multi-channel muscle activation patterns.
	
	\item This paper demonstrates the feasibility of continuous wrist EEG-to-iEMG decoding and shows that matching neural representation capability to motor-task complexity is critical for improving fine-grained brain--muscle interface performance.
	
\end{enumerate}

\section{Related Work}

\subsection{EEG--EMG Corticomuscular Coupling}

The relationship between cortical activity and muscle activation has been widely investigated using EEG--EMG corticomuscular coherence (CMC). CMC quantifies frequency-domain synchronization between motor cortical activity and peripheral muscle activity and provides a non-invasive measure of corticospinal interaction~\cite{mima1999coherence,liu2019cmc}. Significant coupling is commonly observed in the beta band during sustained voluntary contraction, while dynamic force regulation may involve broader frequency ranges~\cite{kristeva2007beta,chakarov2009beta}. These findings indicate that EEG contains task-dependent information related to muscle activation.

CMC, however, describes statistical coupling rather than a direct neural-to-muscular mapping. It cannot reconstruct the continuous temporal evolution or activation intensity of individual muscles. EEG-to-EMG decoding therefore requires models capable of learning nonlinear and time-varying relationships between multichannel cortical signals and muscle activity.

\subsection{Continuous EMG Estimation From EEG}

Continuous EMG estimation from non-invasive scalp EEG remains less explored than EEG-based movement classification. Yoshimura et al. reconstructed the activity of two agonist--antagonist wrist muscles from cortical current sources estimated using a hierarchical Bayesian EEG inverse method during isometric wrist flexion and extension. Their study established the feasibility of recovering individual wrist-muscle activity from non-invasive brain signals~\cite{yoshimura2012reconstruction}. However, it focused on two muscles under static force conditions and relied on EEG source estimation. In contrast, continuous multi-channel iEMG decoding directly from scalp EEG during dynamic wrist movements, where muscle activation patterns are more heterogeneous and weakly correlated, remains insufficiently explored. Their study demonstrated the feasibility of extracting wrist muscle information from cortical signals, but relied on source estimation and mechanically constrained isometric contractions.

Choi extended cortical-source-based EMG reconstruction to continuous upper-limb reaching movements~\cite{choi2013reconstructing}. Sparse linear regression was used to estimate the activity of multiple arm muscles from reconstructed cortical currents. Although this method supported multi-muscle decoding, it depended on computationally intensive source localisation and a linear neural-to-muscular mapping. Liang et al. estimated shoulder muscle activation using a virtual flexor--extensor representation, reducing the dimensionality of the output but discarding the independent activity of individual muscles~\cite{liang2020eeg}.

Deep learning has recently been introduced into EEG-to-EMG decoding. Amiri et al. constructed multi-band spatial EEG representations and employed a CNN--LSTM model to estimate muscle activity during grasp-and-lift movements~\cite{amiri2025decoding}. This approach demonstrated the value of nonlinear spatial--temporal feature extraction but was evaluated on a task characterised by relatively high inter-muscle activation similarity.

Existing studies have established the feasibility of EEG-based EMG estimation, yet dynamic wrist movement remains insufficiently investigated. Previous wrist-related reconstruction relied on cortical source estimation and constrained isometric tasks, whereas recent deep models have mainly focused on grasp-related or proximal upper-limb movements. To the best of our knowledge, this paper presents the first deep learning framework for continuous multi-channel EMG decoding directly from scalp EEG during dynamic wrist flexion and extension.

\subsection{Deep Representation Learning for EEG Decoding}

Deep learning has shown strong capability in extracting nonlinear representations from noisy and non-stationary EEG signals. Convolutional neural networks can learn local temporal and spatial patterns, while compact architectures such as EEGNet use temporal, depthwise, and separable convolutions to capture frequency-specific and channel-dependent information~\cite{schirrmeister2017deep,lawhern2018eegnet}. Recurrent networks and LSTM models are effective for sequential neural signals, whereas Transformer-based models use self-attention to capture long-range temporal dependencies~\cite{hameed2024transformer}. Channel attention and electrode-topology representations further improve spatial modelling by considering unequal channel contributions and scalp electrode arrangement.

These methods demonstrate that spectral, spatial, and temporal EEG features can be learned effectively. However, existing EEG-to-EMG models generally exploit only part of this information. Flattened channel sequences weaken cortical spatial relationships, limited spectral representations may omit complementary motor rhythms, and point-wise regression losses often produce over-smoothed EMG predictions.

Dynamic wrist movement requires the joint modelling of multi-band cortical activity, wrist-related sensorimotor topology, long-range neural--muscular dependencies, and weakly correlated muscle outputs. ST-Topo GAN is developed to integrate these complementary properties within a unified decoding framework matched to wrist neuromuscular complexity.

\section{Methods}

\subsection{Experimental Design}

The experiments comprised three analyses: cross-task comparison of neuromuscular coordination and conventional decoding performance, evaluation of ST-Topo GAN on wrist EEG-to-iEMG decoding, and ablation analysis. WAY-EEG-GAL was used as the reference dataset for cross-task comparison, while model evaluation and ablation analysis were performed on the self-collected wrist dataset.

\begin{figure*}[htbp]
	\centering
	\includegraphics[width=\linewidth]{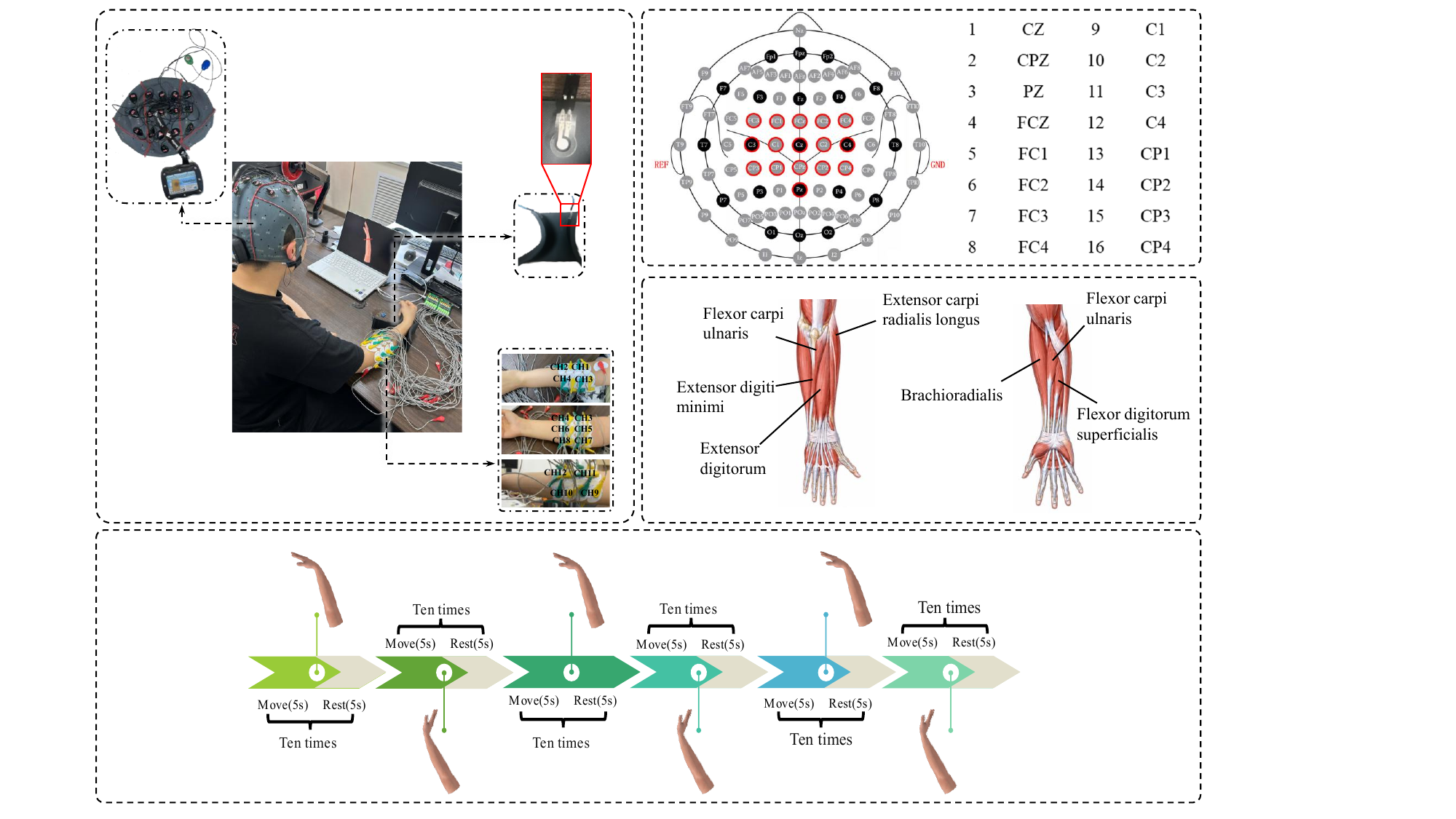}
	\caption{Experimental paradigm of the self-collected wrist movement dataset. Participants performed wrist flexion and extension movements following visual cues. Each trial consisted of a 5 s movement period and a 5 s rest period, with repeated trials conducted across three experimental blocks.}
	\label{fig:wrist_paradigm}
\end{figure*}

\subsection{Datasets and Experimental Paradigms}

\subsubsection{Self-Collected Wrist Movement Dataset}

Ten healthy right-handed volunteers aged $22\pm2$ years performed right-wrist flexion and extension. The forearm was fixed with the palm facing left, and a customised 3D-printed fixture constrained the movement range to maintain consistent wrist amplitudes across trials.

Each participant completed three sessions, each containing three blocks of ten flexion and ten extension trials, yielding 60 trials per session and 180 trials in total. Each trial comprised 5~s of movement followed by 5~s of rest, with 10-s breaks between blocks and 10-min rests between sessions. The paradigm is shown in Fig.~\ref{fig:wrist_paradigm}.

EEG was recorded using a 16-channel g.Nautilus wireless system (g.tec Medical Engineering GmbH, Austria) at 500~Hz, with electrodes covering frontal, central, and parietal sensorimotor regions. Simultaneously, 12-channel surface EMG was recorded from seven wrist- and finger-related forearm muscles at 1000~Hz. The recording configuration is shown in Fig.~\ref{fig:wrist_paradigm}.

EEG quality, electrode impedance, and EMG contact were monitored throughout acquisition. 

\subsubsection{WAY-EEG-GAL Reference Dataset}

The publicly available WAY-EEG-GAL dataset contains synchronised EEG and EMG recordings from twelve healthy participants performing repeated grasp-and-lift movements. EEG was recorded from 32 scalp electrodes at 500~Hz, and five-channel EMG was acquired at 4000~Hz. The dataset served as the reference for inter-muscle coordination and cross-task decoding comparisons.

\subsection{Signal Preprocessing and Sample Construction}

\subsubsection{EEG and EMG Preprocessing}

A unified sampling rate of 1000~Hz was used for EEG--EMG alignment. For the self-collected dataset, EEG was resampled from 500 to 1000~Hz, re-referenced using common average referencing, and filtered with a fourth-order zero-phase Butterworth band-pass filter (0.5--50~Hz) and a 50-Hz notch filter~\cite{widmann2015digital}. EMG was band-pass filtered from 20 to 450~Hz and notch filtered at 50~Hz.

For WAY-EEG-GAL, EEG was resampled from 500 to 1000~Hz, re-referenced using common average referencing, and filtered from 0.1 to 50~Hz with a 50-Hz notch filter. EMG was anti-alias filtered, band-pass filtered from 20 to 450~Hz, and downsampled from 4000 to 1000~Hz~\cite{luciw2014multichannel}.

The preprocessed EEG signals were decomposed into five frequency bands using fourth-order zero-phase Butterworth filters: Delta (0.5--4~Hz), Theta (4--8~Hz), Alpha (8--13~Hz), Beta (13--30~Hz), and Gamma (30--50~Hz)~\cite{newson2019eeg}. The rectified representation of the $c$-th channel in band $f$ was defined as

\begin{equation}
	E_{c,f}(t)
	=
	\left|
	\mathrm{BP}_{f}
	\left(
	\mathrm{EEG}_{c}(t)
	\right)
	\right|,
\end{equation}

where $\mathrm{BP}_{f}(\cdot)$ denotes band-pass filtering in frequency band $f$.

\subsubsection{Dataset Partitioning and Sliding-Window Construction}

Dataset partitioning was performed before sliding-window generation to prevent overlap between training and testing samples. For the self-collected dataset, trials were randomly divided into training and testing sets at a 9:1 ratio using a fixed random seed of 42. For WAY-EEG-GAL, the final recording session was held out for testing.

Sliding windows of 500~ms with a 50-ms step were then generated independently within each subset, with EEG and EMG segments temporally aligned. Each EEG sample was represented as

\begin{equation}
	X
	\in
	\mathbb{R}^{B\times C_f\times C_e\times T},
\end{equation}

where $B$ is the batch size, $C_f=5$ is the number of frequency bands, $C_e$ is the number of EEG channels, and $T=500$ is the temporal length. Here, $C_e=16$ for the wrist dataset and $C_e=32$ for WAY-EEG-GAL.

\subsubsection{Windowed iEMG Activation Representation}

\begin{figure}[htbp]
	\centering
	\includegraphics[width=\linewidth]{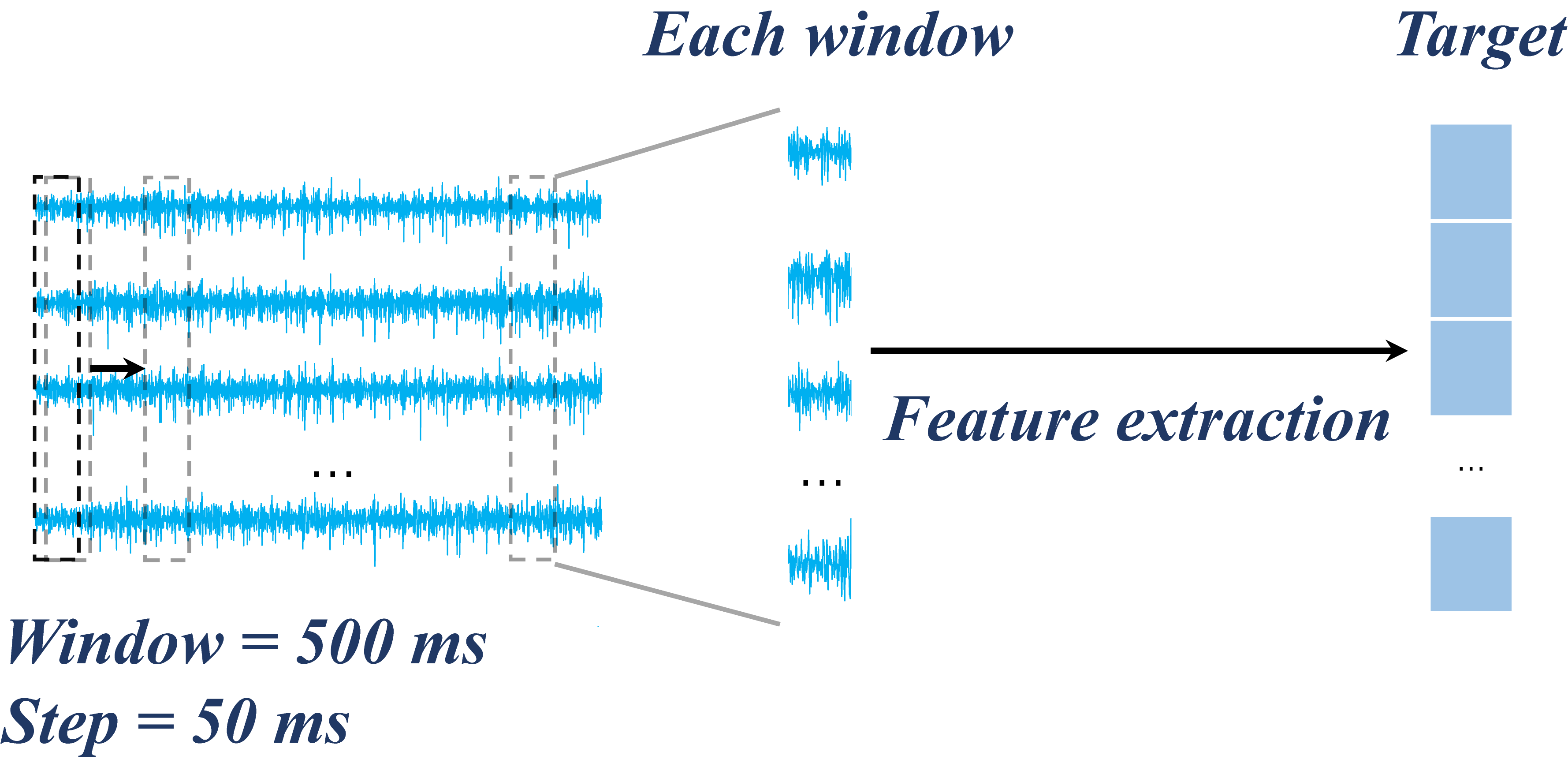}
	\caption{Illustration of EMG preprocessing and windowed muscle activation extraction. Each EMG channel was segmented into 500-ms windows, full-wave rectified, and summed to obtain the continuous iEMG activation target.}
	\label{fig:emg_activation}
\end{figure}

Windowed integrated electromyography (iEMG) was used as the continuous muscle activation target (Fig.~\ref{fig:emg_activation}). Each EMG channel was segmented into 500-ms windows, full-wave rectified, and summed:

\begin{equation}
	\mathrm{iEMG}_{m}(k)
	=
	\sum_{i=1}^{N}
	\left|
	\mathrm{EMG}_{m}(k,i)
	\right|,
\end{equation}

where $m$ denotes the EMG channel, $k$ the temporal window, and $N$ the number of samples within the window. The resulting iEMG values were used as continuous regression targets.

The target tensor was defined as

\begin{equation}
	Y
	\in
	\mathbb{R}^{B\times C_m},
\end{equation}

where $C_m=12$ for the wrist dataset and $C_m=5$ for WAY-EEG-GAL.

\subsubsection{Feature Normalisation}

EEG features were standardised independently for each frequency band using training-set statistics:

\begin{equation}
	\widetilde{X}_{f}
	=
	\frac{
		X_f-\mu_{f}^{\mathrm{tr}}
	}{
		\sigma_{f}^{\mathrm{tr}}+\epsilon
	},
\end{equation}

where $\mu_{f}^{\mathrm{tr}}$ and $\sigma_{f}^{\mathrm{tr}}$ denote the training-set mean and standard deviation of frequency band $f$.

For the wrist dataset, the twelve-channel iEMG targets were globally min--max normalised using training-set extrema:

\begin{equation}
	\widetilde{Y}
	=
	\frac{
		Y-Y_{\min}^{\mathrm{tr}}
	}{
		Y_{\max}^{\mathrm{tr}}-Y_{\min}^{\mathrm{tr}}+\epsilon
	}.
\end{equation}

For WAY-EEG-GAL, MAV targets were independently min--max normalised for each EMG channel using the corresponding training-set statistics. The same normalisation parameters were applied to the test set to prevent data leakage.

\subsection{Neuromuscular Coordination Characterisation}

Inter-muscle correlation was used to characterise neuromuscular coordination. For each participant, Pearson correlations were calculated between all EMG-channel pairs, and the participant-level mean was defined as

\begin{equation}
	\overline{\rho}
	=
	\frac{2}{C_m(C_m-1)}
	\sum_{i=1}^{C_m-1}
	\sum_{j=i+1}^{C_m}
	\rho_{ij},
\end{equation}

where $C_m$ is the number of EMG channels and $\rho_{ij}$ is the Pearson correlation between channels $i$ and $j$. Lower $\overline{\rho}$ indicates weaker inter-muscle activation similarity and more independent recruitment. The participant-level values were compared using a two-sided independent-samples $t$-test ($p<0.05$).

\subsection{Proposed ST-Topo GAN}

\subsubsection{Framework Overview}

The overall architecture of ST-Topo GAN is shown in Fig.~\ref{fig:overview}. The framework estimates twelve-channel wrist iEMG from multi-band EEG using a spectral--spatial topology module, generator $G$, and conditional discriminator $D$.

Given an EEG window $X$, the generator produces $\hat{Y}=G(X)$, while the discriminator receives the EEG condition together with either real or generated iEMG to distinguish real from generated EEG--iEMG pairs. This conditional adversarial constraint promotes iEMG outputs consistent with both the EEG input and the distribution of recorded muscle activation.

\subsubsection{Spectral--Spatial Topological Representation}

The five-band signals from the 16 EEG channels were projected onto a $4\times5$ electrode topology according to their scalp locations, with non-electrode positions set to zero. The resulting tensor is

\begin{equation}
	X_{\mathrm{topo}}
	\in
	\mathbb{R}^{B\times T\times C_f\times H\times W},
\end{equation}

where $C_f=5$, $H=4$, and $W=5$. This representation preserves the spatial relationships among neighbouring sensorimotor electrodes for subsequent feature extraction.

\subsubsection{Temporal Patch Encoding}

The 500-point feature sequence was divided into ten non-overlapping temporal patches, each containing 50 temporal points. The feature vectors within each patch were concatenated and projected into a 128-dimensional embedding space:

\begin{equation}
	z_p
	=
	W_e
	\mathrm{vec}
	\left(
	F_{(p-1)L+1:pL}
	\right)
	+b_e,
\end{equation}

where $L=50$ is the patch length and $p\in\{1,\ldots,10\}$.

Learnable positional embeddings were added to preserve temporal order:

\begin{equation}
	Z_0=[z_1,z_2,\ldots,z_{10}]+E_{\mathrm{pos}}.
\end{equation}

The embedded sequence was processed by three Transformer blocks. Each block contained eight-head self-attention, a 512-dimensional feed-forward network, residual connections, layer normalisation, and a dropout rate of 0.3:

\begin{equation}
	Z_l'
	=
	\mathrm{LN}
	\left(
	Z_{l-1}
	+
	\mathrm{MHSA}(Z_{l-1})
	\right),
\end{equation}

\begin{equation}
	Z_l
	=
	\mathrm{LN}
	\left(
	Z_l'
	+
	\mathrm{FFN}(Z_l')
	\right).
\end{equation}

Temporal self-attention enables the model to associate muscle activation with cortical information distributed across movement preparation, activation onset, sustained contraction, and feedback regulation within the observation window.

The encoded tokens were flattened and mapped through a 256-unit fully connected layer with LeakyReLU and dropout. A sigmoid output layer generated the twelve-channel normalised muscle activation:

\begin{equation}
	\hat{Y}
	=
	G(X)
	\in
	\mathbb{R}^{B\times12}.
\end{equation}

\begin{figure*}[t]
	\centering
	\includegraphics[width=\linewidth]{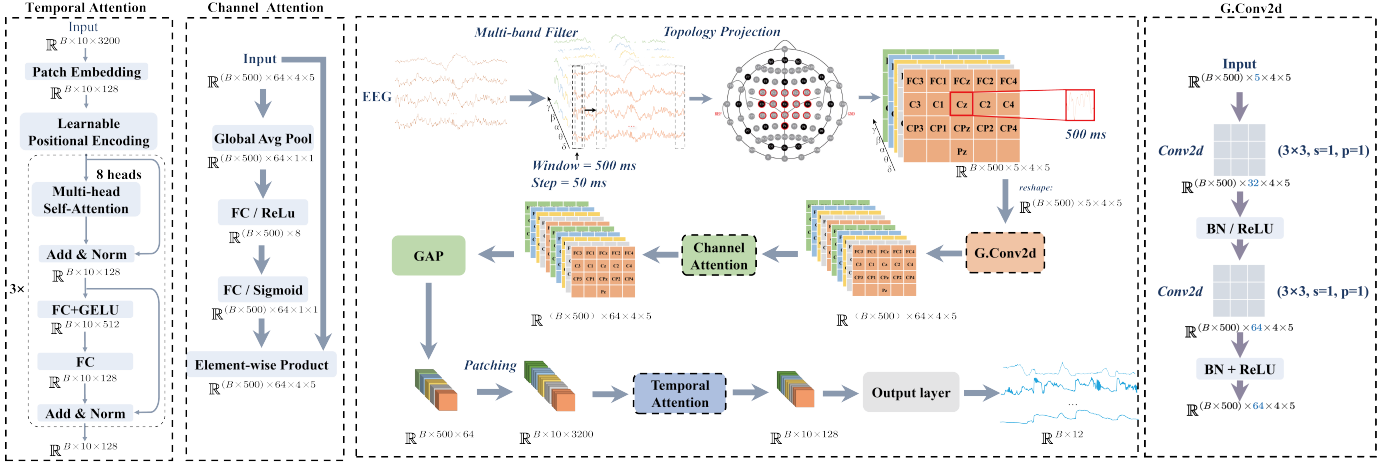}
	\caption{Architecture of the proposed wrist-oriented ST-Topo GAN generator. Five-band EEG signals are mapped onto a $4\times5$ electrode topology, encoded using spatial convolution and channel attention, and divided into temporal patches for self-attention-based modelling.}
	\label{fig:generator_architecture}
\end{figure*}

The generator architecture is shown in Fig.~\ref{fig:generator_architecture}.

\subsubsection{Conditional Discriminator}

The discriminator jointly processes the EEG condition and the corresponding iEMG vector to distinguish recorded EEG--iEMG pairs from generated ones. Its EEG branch adopts the same $4\times5$ spectral--spatial topology used by the generator.

At each temporal point, spatial features are extracted using two two-dimensional convolutional layers with $3\times3$ kernels and 16 and 32 output channels, respectively. The resulting $32\times4\times5$ feature map is flattened into a 640-dimensional vector, yielding a temporal sequence of spatial EEG representations.

Temporal dependencies are then modelled using two one-dimensional convolutional layers. The first maps the 640-dimensional input sequence to 128 channels, and the second maps 128 channels to 64 channels. Both layers use a kernel size of 15, a stride of 4, and padding of 7.

\begin{figure}[t]
	\centering
	\includegraphics[width=\linewidth]{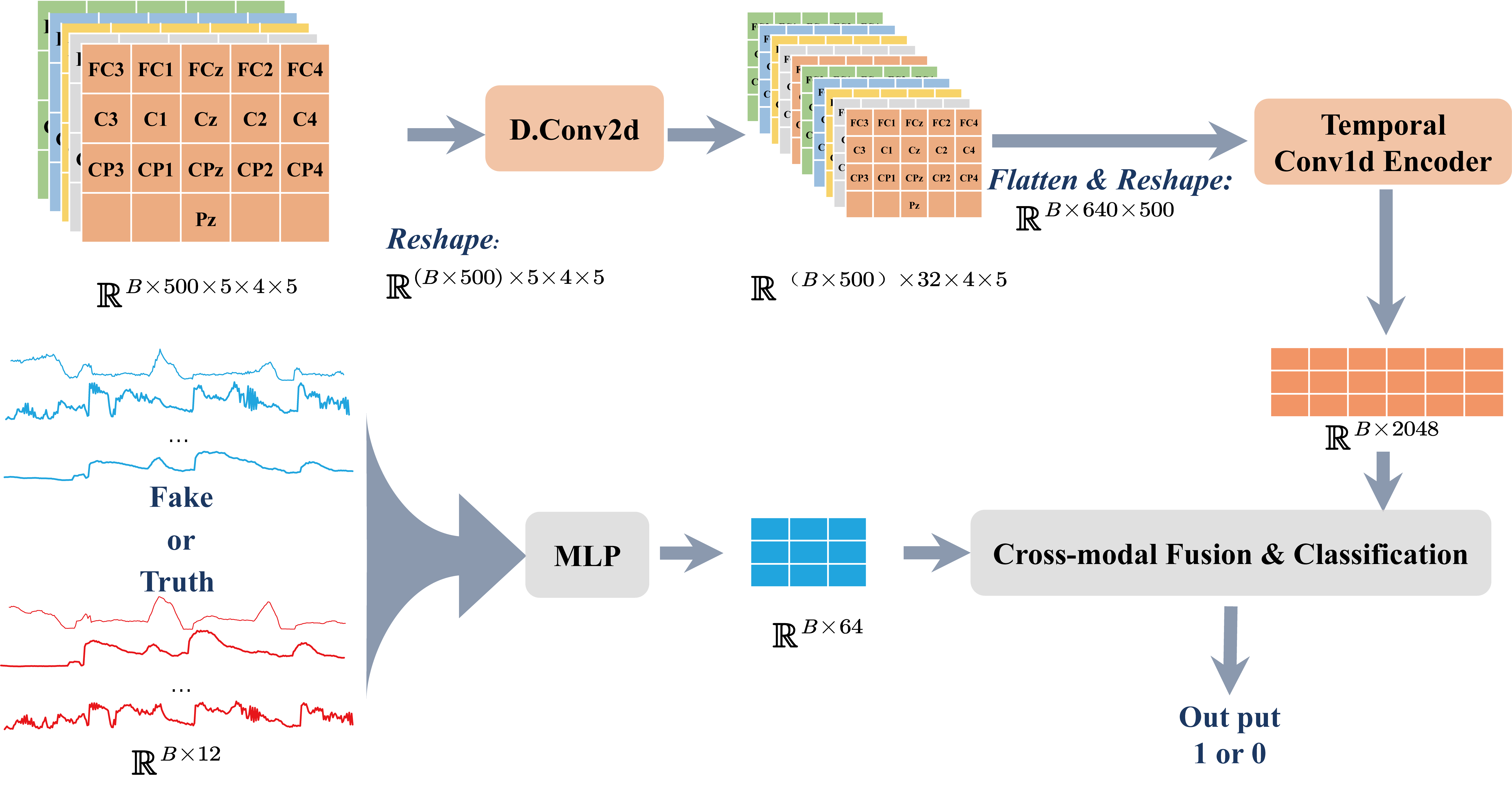}
	\caption{Conditional adversarial learning framework. The discriminator distinguishes real EEG--EMG pairs from EEG--generated-EMG pairs and provides distribution-level feedback to the generator.}
	\label{fig:adversarial_framework}
\end{figure}

The twelve-channel iEMG vector is independently projected into a 64-dimensional representation:

\begin{equation}
	H_{\mathrm{EMG}}
	=
	\phi
	\left(
	W_yY+b_y
	\right),
\end{equation}

where $\phi(\cdot)$ denotes LeakyReLU activation. The EEG and EMG features are concatenated and processed by a 256-unit fully connected layer and a sigmoid classifier:

\begin{equation}
	D(X,Y)
	=
	\sigma
	\left[
	W_d
	\left(
	H_{\mathrm{EEG}}
	\Vert
	H_{\mathrm{EMG}}
	\right)
	+b_d
	\right],
\end{equation}

where $\Vert$ denotes feature concatenation. Conditioning the discriminator on EEG encourages the generated EMG to reflect both the distribution of real wrist muscle activation and its correspondence with the cortical input. The adversarial framework is illustrated in Fig.~\ref{fig:adversarial_framework}.

\subsubsection{Learning Objectives}

The discriminator was trained using binary cross-entropy to distinguish real EEG--EMG pairs from generated pairs:

\begin{equation}
	\mathcal{L}_{D}=\frac{1}{2}\operatorname{BCE}\!\left(D(X,Y),1\right)+\frac{1}{2}\operatorname{BCE}\!\left(D(X,G(X)),0\right).
\end{equation}

The generator combined adversarial learning with cross-muscle correlation and point-wise reconstruction constraints. The adversarial loss was defined as $\mathcal{L}_{\mathrm{adv}}=\operatorname{BCE}(D(X,G(X)),1)$.

To preserve the relative recruitment pattern across EMG channels, the correlation coefficient for sample $b$ was calculated as

\begin{equation}
	\rho_b=\frac{\sum_{m=1}^{C_m}\tilde{\hat{y}}_{b,m}\tilde{y}_{b,m}}{\sqrt{\sum_{m=1}^{C_m}\tilde{\hat{y}}_{b,m}^{2}}\sqrt{\sum_{m=1}^{C_m}\tilde{y}_{b,m}^{2}}+\epsilon},
\end{equation}

where $\tilde{\hat{y}}_{b,m}=\hat{y}_{b,m}-\overline{\hat{y}}_b$ and $\tilde{y}_{b,m}=y_{b,m}-\overline{y}_b$. The corresponding cross-muscle correlation loss was

\begin{equation}
	\mathcal{L}_{\mathrm{cc}}=1-\frac{1}{B}\sum_{b=1}^{B}\rho_b.
\end{equation}

The point-wise reconstruction terms were defined as $\mathcal{L}_{\mathrm{L1}}=\|\hat{Y}-Y\|_1/(BC_m)$ and $\mathcal{L}_{\mathrm{logcosh}}=\operatorname{mean}[\log\cosh(\hat{Y}-Y)]$. The complete generator objective was

\begin{equation}
	\mathcal{L}_{G}=0.1\mathcal{L}_{\mathrm{adv}}+50\mathcal{L}_{\mathrm{cc}}+20\mathcal{L}_{\mathrm{L1}}+50\mathcal{L}_{\mathrm{logcosh}}.
\end{equation}

The correlation and reconstruction terms constrain multi-channel recruitment patterns and point-wise activation errors, respectively, while adversarial learning provides a complementary conditional distribution constraint.

\subsection{Baseline Models}

Five representative EEG decoding architectures were included as baselines: 1D-CNN, LSTM, CNN-LSTM, EEGNet and Transformer.

The 1D-CNN extracted hierarchical local temporal features using stacked one-dimensional convolutional blocks. The LSTM model captured sequential dependencies directly from multichannel EEG. CNN-LSTM combined local convolutional feature extraction with recurrent temporal modelling. EEGNet employed temporal, depthwise, and separable convolutions to construct a compact EEG representation. The Transformer used multi-head self-attention to model long-range temporal interactions.

All models used identical trial partitions, EEG frequency bands, window lengths, step sizes, muscle-activation targets, and evaluation procedures. Input tensors were rearranged according to the requirements of each architecture. Independent subject-dependent models were trained for every participant.

\subsection{Experimental Protocols}

\subsubsection{Cross-Task Complexity Comparison}

The cross-task experiment evaluated whether wrist movement imposed greater demands on conventional EEG-to-EMG decoding than the WAY-EEG-GAL grasp-and-lift task. Inter-muscle correlations and baseline decoding performance were compared between the two datasets using their respective subject-dependent data partitions. ST-Topo GAN was excluded from this experiment because the public dataset was used only to establish a reference for task-related decoding difficulty. The relative PCC variation was calculated as

\begin{equation}
	\Delta_{\mathrm{PCC}}=\frac{\mathrm{PCC}_{\mathrm{wrist}}-\mathrm{PCC}_{\mathrm{GAL}}}{\mathrm{PCC}_{\mathrm{GAL}}}\times100\%,
\end{equation}

where a negative value indicates lower decoding performance on the wrist dataset.

\subsubsection{Wrist EEG-to-EMG Decoding}

The principal model comparison was conducted on the self-collected wrist movement dataset. ST-Topo GAN and the baseline models were evaluated using identical participant-specific 9:1 trial-level partitions. Performance was calculated independently for each of the twelve EMG channels, averaged to obtain a participant-level score, and subsequently summarised across the ten participants.

\subsubsection{Ablation Analysis}

The contribution of the proposed components was examined using four ablation variants: \textit{NoBands}, in which the multi-band representation was removed; \textit{NoTopo}, in which the electrode topology was removed; \textit{NoAttention}, in which temporal self-attention was removed; and \textit{NoGAN}, in which the discriminator and adversarial loss were removed. All variants used identical data partitions, preprocessing procedures, and training targets. The ablation analysis was conducted on three representative participants (\textit{czr}, \textit{lyt}, and \textit{nst}).

\subsection{Implementation and Training Configuration}

All models were implemented in PyTorch and trained in a subject-dependent manner. ST-Topo GAN was trained for 100 epochs with a batch size of 64. The generator and discriminator were optimised using Adam with a learning rate of $2\times10^{-4}$ and parameters $\beta_1=0.5$ and $\beta_2=0.999$. One discriminator update and one generator update were performed for each mini-batch. The number of training epochs was selected through preliminary hyperparameter optimisation, and the final-epoch model was used for testing. A fixed random seed of 42 was used to reproduce the wrist trial partition.

\subsection{Evaluation and Statistical Analysis}

Decoding performance was assessed using the Pearson correlation coefficient (PCC) and normalised root mean square error (nRMSE). PCC was treated as the primary metric because it evaluates the temporal correspondence between predicted and reference muscle activation, whereas nRMSE was used as a complementary measure of amplitude reconstruction error. Both metrics were calculated independently for each EMG channel and averaged across channels to obtain a participant-level result. Group results are reported as the mean and standard deviation across participants.

Differences in average inter-muscle correlation between the two datasets were assessed using a two-sided independent-samples $t$-test. Model differences on the wrist dataset were evaluated using the Friedman test, followed by pairwise Wilcoxon signed-rank tests between ST-Topo GAN and each baseline model. Holm correction was applied to account for multiple comparisons, and statistical significance was defined as $p<0.05$.

\section{Results}

\subsection{Cross-task Complexity Comparison of Wrist EEG-to-iEMG Decoding}

The difference in neuromuscular coordination between grasp-and-lift and wrist movements was first investigated by comparing inter-muscle correlations. The WAY-EEG-GAL dataset exhibited an average inter-muscle correlation of $0.6496\pm0.1294$, whereas the self-collected wrist movement dataset showed a lower correlation of $0.4662\pm0.0738$. The difference between the two datasets was statistically significant ($t=3.790$, $p=0.0011$).

\begin{table}[t]
	\caption{Average inter-muscle correlations between the WAY-EEG-GAL dataset and the self-collected wrist movement dataset.}
	\label{tab:muscle_corr}
	\centering
	\begin{tabular}{lcc}
		\hline
		Dataset & Mean correlation & Std. \\
		\hline
		WAY-EEG-GAL & 0.6496 & 0.1294 \\
		Wrist movement dataset & 0.4662 & 0.0738 \\
		\hline
	\end{tabular}
\end{table}

\begin{figure}[t]
	\centering
	\includegraphics[width=\linewidth]{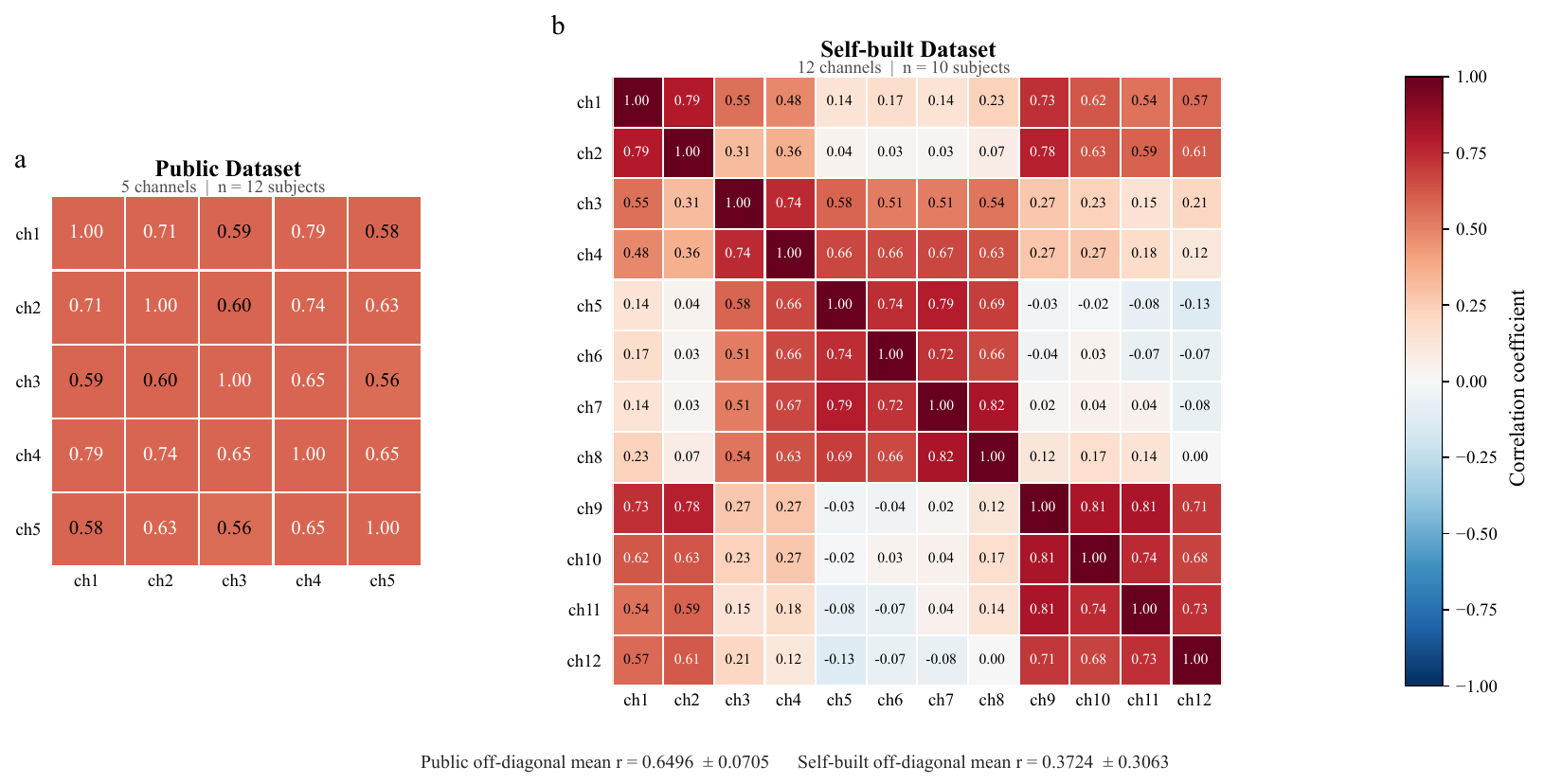}
	\caption{Inter-channel Pearson correlation matrices of EMG activation in the WAY-EEG-GAL dataset and the self-collected wrist movement dataset. The wrist dataset exhibits weaker and less uniform inter-muscle correlations.}
	\label{fig:emg_correlation}
\end{figure}

The inter-channel Pearson correlation matrices of EMG activation are illustrated in Fig.~\ref{fig:emg_correlation}. Compared with grasp-and-lift movements, wrist movements exhibited weaker and more heterogeneous correlations among EMG channels.

\begin{figure}[t]
	\centering
	\includegraphics[width=\linewidth]{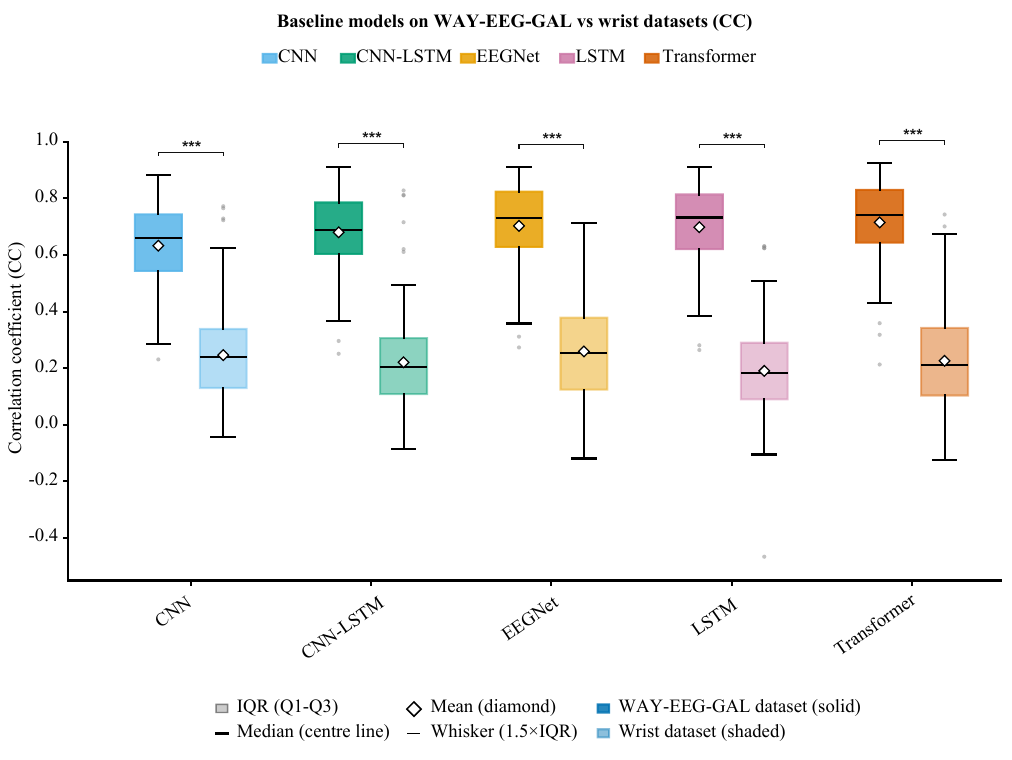}
	\caption{Cross-task decoding performance comparison of five baseline EEG-to-iEMG models on the WAY-EEG-GAL dataset and the self-collected wrist movement dataset. All baseline models exhibit performance degradation when applied to wrist movements.}
	\label{fig:cross_task}
\end{figure}

To further evaluate whether conventional EEG-to-iEMG decoding models are affected by motor-task characteristics, five representative baseline architectures were trained and tested on both the WAY-EEG-GAL dataset and the self-collected wrist movement dataset. The participant-level PCC values and relative performance variations between the two tasks are summarised in Table~\ref{tab:cross_task_baselines}, and the overall comparison is shown in Fig.~\ref{fig:cross_task}.

\begin{table*}[t]
	\caption{Cross-task decoding performance of five baseline EEG-to-iEMG models on the WAY-EEG-GAL dataset and the wrist movement dataset. Results are reported as mean $\pm$ standard deviation across participants.}
	\label{tab:cross_task_baselines}
	\centering
	\begin{tabular}{lccc}
		\hline
		Model & WAY-EEG-GAL & Wrist dataset & Variation (\%) \\
		\hline
		
		1D-CNN
		& $0.6330\pm0.1024$
		& $0.2463\pm0.1071$
		& $-61.1$ \\
		
		CNN-LSTM
		& $0.6803\pm0.0837$
		& $0.2209\pm0.1194$
		& $-67.5$ \\
		
		EEGNet
		& $0.7031\pm0.0753$
		& $0.2592\pm0.0987$
		& $-63.1$ \\
		
		LSTM
		& $0.6983\pm0.0814$
		& $0.1905\pm0.0873$
		& $-72.7$ \\
		
		Transformer
		& $\mathbf{0.7152\pm0.0829}$
		& $0.2262\pm0.1069$
		& $-68.4$ \\
		
		\hline
	\end{tabular}
\end{table*}

As shown in Fig.~\ref{fig:cross_task}, all baseline models achieved lower PCC values on the wrist movement dataset compared with the WAY-EEG-GAL dataset. The performance reduction ranged from 61.1\% for 1D-CNN to 72.7\% for LSTM. Although different architectures exhibited different absolute performance levels across datasets, convolutional, recurrent, hybrid, and attention-based models all showed consistent degradation during wrist movement decoding.

Together with the reduced inter-muscle correlations shown in Fig.~\ref{fig:emg_correlation}, these results indicate that wrist movements introduce additional challenges for continuous EEG-to-iEMG decoding compared with conventional grasp-and-lift tasks.

\subsection{Wrist EEG-to-EMG Decoding Performance}

ST-Topo GAN was compared with the five conventional decoding models on the self-collected wrist movement dataset. PCC was used as the primary metric for evaluating the temporal consistency between the predicted and reference muscle activation sequences, while nRMSE was reported as a complementary measure of point-wise amplitude error. The results are summarised in Table~\ref{tab:wrist_model_compare}.

\begin{table}[t]
	\caption{EEG-to-EMG decoding performance on the self-collected wrist movement dataset. Results are reported as mean $\pm$ standard deviation across participants. The best result for each metric is shown in bold.}
	\label{tab:wrist_model_compare}
	\centering
	\begin{tabular}{lcc}
		\hline
		Model & PCC & nRMSE \\
		\hline
		1D-CNN
		& $0.2463\pm0.1071$
		& $0.2072\pm0.0281$ \\
		
		CNN-LSTM
		& $0.2209\pm0.1194$
		& $0.2311\pm0.0387$ \\
		
		EEGNet
		& $0.2592\pm0.0987$
		& $\mathbf{0.2026\pm0.0269}$ \\
		
		LSTM
		& $0.1905\pm0.0873$
		& $0.2167\pm0.0346$ \\
		
		Transformer
		& $0.2262\pm0.1069$
		& $0.2108\pm0.0238$ \\
		
		ST-Topo GAN
		& $\mathbf{0.4436\pm0.0981}$
		& $0.2314\pm0.0310$ \\
		\hline
	\end{tabular}
\end{table}

ST-Topo GAN achieved the highest PCC of $0.4436\pm0.0981$. EEGNet was the strongest conventional baseline,
with a PCC of $0.2592\pm0.0987$. The proposed model therefore increased the average PCC by 0.1844, corresponding to a relative improvement of 71.1\% over the strongest baseline.

\begin{figure}[t]
	\centering
	\includegraphics[width=\linewidth]{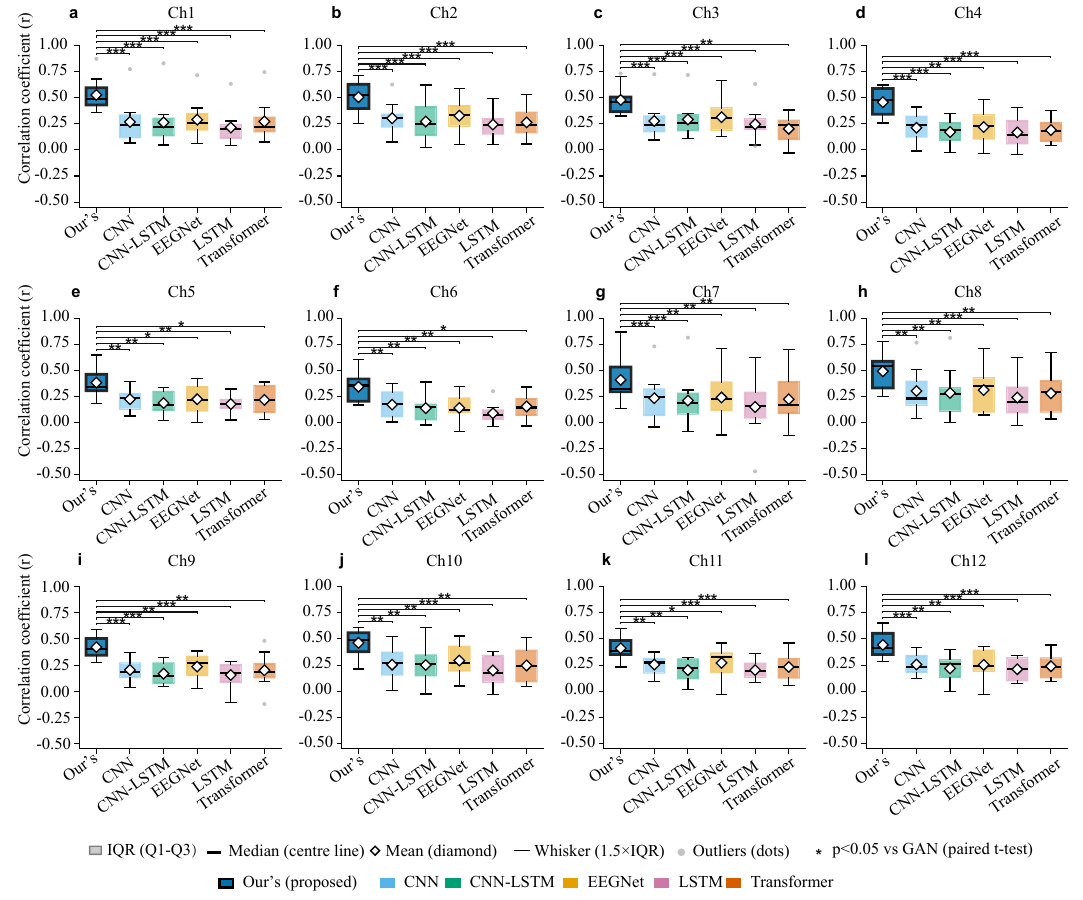}
	\caption{PCC comparison of ST-Topo GAN and representative baseline models on the self-collected wrist movement dataset. Each result represents the average performance across twelve EMG channels for each participant.}
	\label{fig:wrist_statistics}
\end{figure}

A Friedman test revealed an overall difference in PCC among the six models ($\chi^2(5)=26.46$, $p<0.001$). Pairwise Wilcoxon signed-rank tests with Holm correction showed that ST-Topo GAN achieved a higher PCC than each baseline model (all adjusted $p=0.0098$).

The nRMSE values ranged from 0.2026 to 0.2314. EEGNet achieved the lowest nRMSE, while the nRMSE of ST-Topo GAN was nearly identical to that of CNN-LSTM. These results show that the principal performance gain of ST-Topo GAN was observed in the temporal correspondence of the reconstructed muscle activation, rather than in the minimum
point-wise amplitude error.

The participant-level performance comparison is illustrated in Fig.~\ref{fig:wrist_statistics}. ST-Topo GAN maintained a higher PCC than the evaluated baseline architectures across the wrist dataset.

\begin{figure}[htbp]
	\centering
	\includegraphics[width=\linewidth]{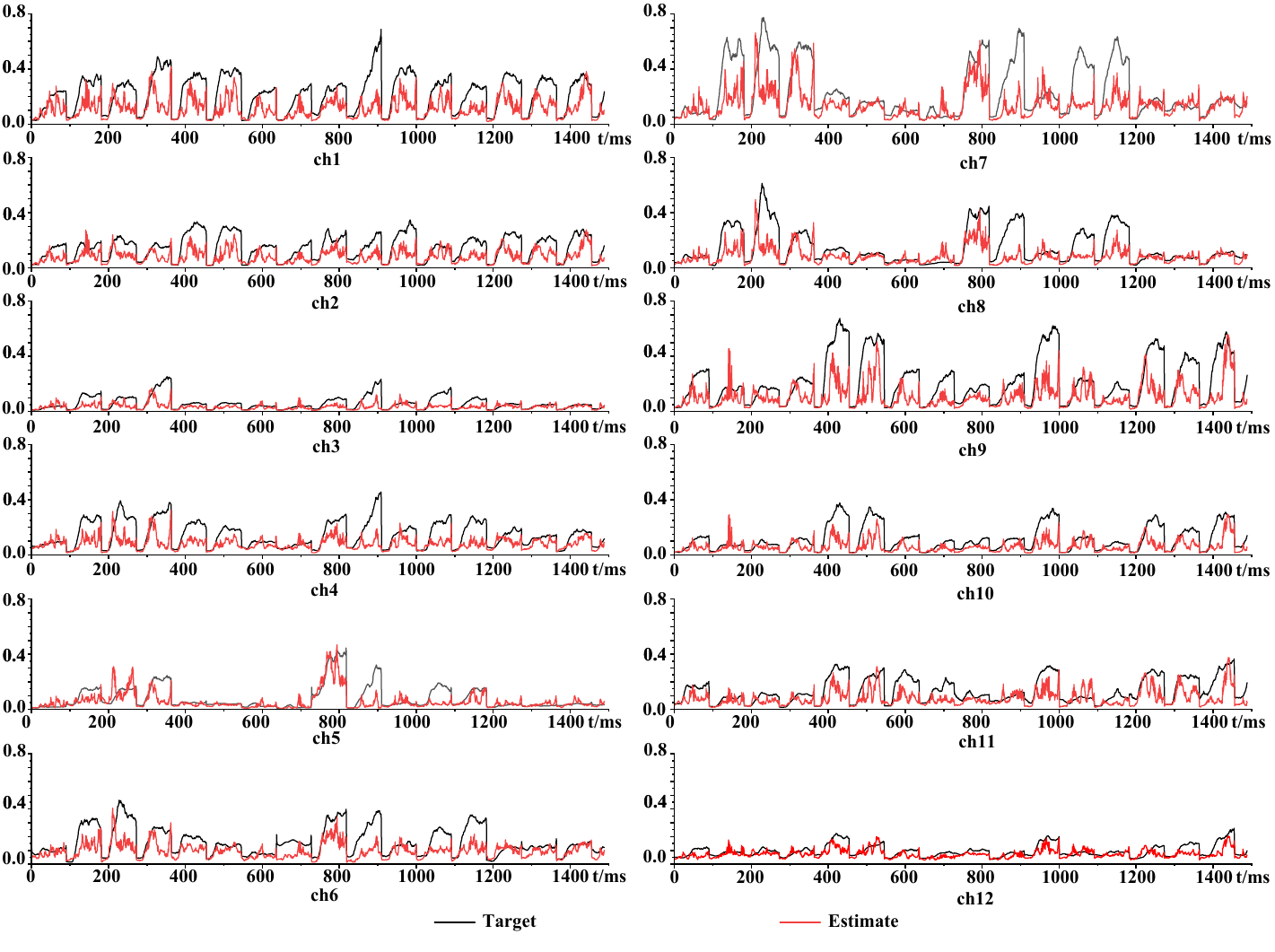}
	\caption{Representative twelve-channel wrist muscle activation reconstruction obtained using ST-Topo GAN. Ground-truth iEMG targets and predicted activation sequences are shown over consecutive testing windows.}
	\label{fig:wrist_waveform}
\end{figure}

Representative reconstruction results are shown in Fig.~\ref{fig:wrist_waveform}. The predicted iEMG sequences followed the principal activation increases and decreases of the reference signals across multiple EMG channels. Deviations remained at several local peaks and low-amplitude intervals.

\subsection{Ablation Study}

Ablation experiments were conducted on three representative participants from the wrist movement dataset. Four variants were constructed by removing the multi-band spectral representation, electrode topology, temporal attention, or adversarial learning. All variants used identical data
partitions and training targets.

\begin{table}[t]
	\caption{Ablation study on three representative participants from the self-collected wrist movement dataset. PCC reduction is calculated relative to the complete ST-Topo GAN.}
	\label{tab:wrist_ablation}
	\centering
	\begin{tabular}{lcc}
		\hline
		Model & PCC & Variation (\%) \\
		\hline
		NoBands
		& $0.4510\pm0.0802$
		& -17.7 \\
		
		NoTopo
		& $0.4701\pm0.0356$
		& -14.2 \\
		
		NoAttention
		& $0.5093\pm0.0347$
		& -7.1 \\
		
		NoGAN
		& $0.5120\pm0.0266$
		& -6.6 \\
		
		ST-Topo GAN
		& $\mathbf{0.5481\pm0.0905}$
		& -- \\
		\hline
	\end{tabular}
\end{table}

The complete ST-Topo GAN achieved the highest PCC of $0.5481\pm0.0905$. Removing the multi-band representation produced the largest performance reduction, decreasing the PCC to $0.4510\pm0.0802$. Removing the spatial topology reduced the PCC to $0.4701\pm0.0356$.

The variants without temporal attention and adversarial learning achieved PCC values of $0.5093\pm0.0347$ and
$0.5120\pm0.0266$. Although these reductions were smaller than those caused by removing the spectral or topological representations, both variants remained below the complete model. The ablation results
indicate that the spectral, spatial, temporal, and adversarial components provide complementary contributions to wrist EEG-to-EMG decoding.

\section{Discussion}

This study proposed ST-Topo GAN for continuous EEG-to-iEMG decoding during dynamic wrist movement. The wrist dataset exhibited lower and more heterogeneous inter-muscle correlations than the grasp-and-lift reference dataset, while conventional convolutional, recurrent, hybrid, and attention-based models all showed substantial reductions in PCC on the wrist task. ST-Topo GAN achieved a significantly higher PCC than the evaluated baselines, supporting the view that wrist EEG-to-iEMG decoding benefits from a representation structure aligned with the spectral, spatial, temporal, and multi-muscle characteristics of wrist control.

The lower inter-muscle correlations observed in the wrist dataset suggest that the recorded forearm muscles were less dominated by a common activation pattern. Pairwise inter-muscle correlations quantify the similarity between muscle activation profiles rather than directly identifying the underlying control mechanisms or muscle synergies. The present results therefore indicate greater heterogeneity among the twelve EMG activation channels during wrist movement. Wrist flexion and extension involve task-dependent activation of flexors, extensors, and stabilising muscles under distinct neural and biomechanical constraints~\cite{davella2003synergies,ting2007neuromechanics}, making the twelve-channel activation target more difficult to reconstruct than activation patterns exhibiting higher and more consistent inter-muscle similarity. The performance reduction observed across all baseline architectures further indicates that temporal feature extraction alone is insufficient for this task. Effective wrist decoding must also preserve task-dependent cortical rhythms, the spatial organisation of sensorimotor activity, and the relative activation patterns across multiple muscles.

The architecture of ST-Topo GAN was designed around these requirements. Motor preparation and execution are accompanied by frequency-specific changes in sensorimotor rhythms, particularly event-related modulation within the alpha and beta ranges~\cite{pfurtscheller1999erd,neuper2001dynamics}. EEG--EMG coupling during voluntary wrist flexion and extension has also been observed within sensorimotor frequency ranges, demonstrating that scalp EEG contains information related to peripheral muscle activity~\cite{halliday1998coupling}. Multi-band decomposition therefore allows the model to retain complementary oscillatory information that may be obscured by a single broadband representation. The electrode topology further preserves the spatial relationships among neighbouring sensorimotor electrodes, thereby enabling the generator to model distributed cortical activity rather than treating EEG channels as independent sequences.

Temporal self-attention integrates information distributed throughout the 500-ms observation window, which is relevant because cortical activity and muscle activation are not related through a purely instantaneous mapping. Previous work has shown that individual wrist flexor and extensor activity can be reconstructed from EEG-derived cortical currents, but also highlighted the difficulty of separating muscle-specific information from mixed scalp recordings~\cite{yoshimura2012reconstruction}. In ST-Topo GAN, temporal modelling is combined with spectral--spatial representation to learn this complex neural-to-muscular relationship directly from scalp EEG.

Conditional adversarial learning complements point-wise reconstruction losses by encouraging the generated output to follow the distributional structure of recorded multi-channel iEMG. Adversarial learning is particularly useful when point-wise objectives alone favour averaged or over-smoothed predictions, because the discriminator provides an additional data-driven constraint on higher-order output characteristics~\cite{creswell2018gan}. The ablation results support the complementary roles of the proposed components. Removing the multi-band representation produced the largest PCC reduction, followed by removal of the electrode topology, whereas temporal attention and adversarial learning provided further improvements. The ablation analysis on three representative participants was intended to compare the relative effects of the architectural components rather than to support population-level statistical inference.

ST-Topo GAN achieved a substantially higher PCC but did not obtain the lowest nRMSE. This difference reflects the complementary properties of the two metrics. PCC measures the correspondence between the temporal variations of the predicted and recorded activation, whereas nRMSE is more sensitive to point-wise amplitude errors. The correlation-based reconstruction loss and adversarial objective emphasise temporal consistency, relative activation patterns, and multi-channel output structure, which may explain why the improvement was more pronounced in PCC. The proposed model is therefore particularly effective at reconstructing the temporal evolution and inter-muscle activation structure of wrist muscle activity, while absolute amplitude calibration remains an area for further improvement.

The higher temporal correspondence achieved by ST-Topo GAN is relevant to continuous brain--muscle interfaces, where activation onset, offset, temporal modulation, and relative muscle activation are important for interpreting motor intention. Earlier EEG-to-EMG studies demonstrated that continuous muscle activity can provide richer control information than discrete movement labels for reproducing natural motor output~\cite{yoshimura2012reconstruction}. Continuous multi-channel iEMG estimates may therefore support wrist rehabilitation robots, neuroprosthetic systems, and multi-channel functional electrical stimulation. However, applications that directly regulate stimulation intensity or actuator force also require accurate amplitude estimation. Future models may combine correlation-oriented learning with channel-specific calibration, adaptive amplitude losses, or subject-specific output normalisation to improve nRMSE without compromising temporal correspondence.

The present findings also highlight the importance of matching decoding architectures to the target motor task. Inter-muscle activation patterns are shaped by both neural control strategies and task-specific biomechanical requirements~\cite{ting2007neuromechanics}. Models that perform well on grasp-and-lift movements characterised by relatively high inter-muscle activation similarity may therefore exhibit reduced performance when applied to wrist movements with more heterogeneous and weakly correlated muscle activation patterns. ST-Topo GAN is not intended as a universally optimal EEG-to-iEMG decoder; rather, its contribution is to demonstrate that task-oriented spectral, spatial, temporal, and output constraints can improve decoding when the model architecture is aligned with the neuromuscular characteristics of the target movement.

\section{Conclusion}

This study proposes ST-Topo GAN for continuous multi-channel EEG-to-iEMG decoding during dynamic wrist movement. Cross-task analysis showed that the wrist dataset exhibited lower and more heterogeneous inter-muscle correlations than the WAY-EEG-GAL grasp-and-lift dataset. Conventional convolutional, recurrent, hybrid, and attention-based models also exhibited consistent performance reductions on the wrist task, demonstrating the increased difficulty of reconstructing heterogeneous wrist-muscle activation patterns from scalp EEG.

To address this challenge, ST-Topo GAN combines multi-band EEG decomposition, sensorimotor electrode topology, temporal self-attention, and conditional adversarial learning within a unified decoding framework. The multi-band and topology-based representations preserve complementary spectral and spatial information, while temporal modelling and adversarial generation support the reconstruction of dynamic multi-channel iEMG activation. On the self-collected wrist dataset, ST-Topo GAN achieved an average PCC of 0.4436 and outperformed all evaluated baseline models. The ablation results further demonstrated the complementary contributions of the spectral, spatial, temporal, and adversarial components.

These findings indicate that continuous wrist EEG-to-iEMG decoding benefits from an architecture aligned with the task-specific characteristics of cortical activity and heterogeneous multi-muscle activation. ST-Topo GAN provides a promising foundation for fine-grained brain--muscle interfaces and may support future applications in wrist rehabilitation, neuroprosthetic control, and functional electrical stimulation.

\section*{Acknowledgment}

The human-subject illustration in Fig.~\ref{fig:overview} was generated using an OpenAI image-generation system and subsequently edited by the authors. The scientific content and technical accuracy of the figure were determined and verified by the authors.

\end{document}